\documentclass{article}

\usepackage{PRIMEarxiv}
\usepackage{cite}
\usepackage{amsmath,amssymb,amsfonts}
\usepackage{algorithmic}
\usepackage{graphicx}
\usepackage{textcomp}
\usepackage{xcolor}
\usepackage[utf8]{inputenc} 
\usepackage[T1]{fontenc}    
\usepackage{url}            
\usepackage{booktabs}       
\usepackage{amsfonts}       
\usepackage{nicefrac}       
\usepackage{microtype}      
\usepackage{lipsum}
\usepackage{fancyhdr}       
\usepackage{graphicx}       
\usepackage{booktabs}
\usepackage{xcolor}
\usepackage[normalem]{ulem}
\usepackage{balance}
\usepackage{authblk}  

\def\BibTeX{{\rm B\kern-.05em{\sc i\kern-.025em b}\kern-.08em
    T\kern-.1667em\lower.7ex\hbox{E}\kern-.125em}}
\begin{document}

\title{A CLIP-based Siamese Approach for Meme Classification 
\thanks{This work is partially funded by the Horizon 2020 European project vera.ai under grant agreement no. 101070093 and AI4Media under grant agreement no. 951911, part of the project PCI2022-134990-2 (MARTINI) of the CHISTERA IV Cofund 2021 program, funded by MCIN/AEI/10.13039/501100011033 and by the “European Union NextGenerationEU/PRTR”, by the Spanish Ministry of Science and Innovation under FightDIS (PID2020-117263GB-100) grant and MCIN/AEI/10.13039/501100011033/ and European Union NextGeneration EU/PRTR for XAI-Disinfodemics (PLEC2021-007681) grant, by European Comission under IBERIFIER Plus - Iberian Digital Media Observatory (DIGITAL-2023-DEPLOY-04-EDMO-HUBS 101158511), and by "Convenio Plurianual with the Universidad Politécnica de Madrid in the actuation line of \textit{Programa de Excelencia para el Profesorado Universitario}". This publication is also part of the I+D+i project PLEC2021-007681, financed by MCIN/AEI/10.13039/501100011033/ and the European Union NextGeneration/PRTR. This work is funded by the National Natural Science Foundation of China (No. 62171114), and the Fundamental Research Funds for the Central Universities (Nos. N232400412 and DUT22RC(3)099)\\
\color{red}\textbf{Warning}: This article tackles subjects that some readers may deem offensive such as misogyny, racism or calls to violence, with explicit examples and commentary on the matter.\color{black}}
}

\author[1]{Javier Huertas-Tato}
\author[2]{Christos Koutlis}
\author[2]{Symeon Papadopoulos}
\author[1]{David Camacho}
\author[2]{Ioannis Kompatsiaris}

\affil[1]{Departamento de Sistemas Informáticos, \textit{Universidad Politécnica de Madrid}, Madrid, Spain.}
\affil[2]{Information Technologies Institute, \textit{Centre for Research and Technology Hellas}, Thessaloniki, Greece.}

\affil[ ]{\texttt{\{javier.huertas.tato, david.camacho\}@upm.es | \{ckoutlis, papadop, ikom@iti\}.gr}}

\newcommand{\href}[2]{#2}

\maketitle

\begin{abstract}
Memes are an increasingly prevalent element of online discourse in social networks, especially among young audiences. They carry ideas and messages that range from humorous to hateful, and are widely consumed. Their potentially high impact requires adequate means of control to moderate their use in large scale. In this work, we propose SimCLIP a deep learning-based architecture for cross-modal understanding of memes, leveraging a pre-trained CLIP encoder to produce context-aware embeddings and a Siamese fusion technique to capture the interactions between text and image. We perform an extensive experimentation on seven meme classification tasks across six datasets. We establish a new state of the art in Memotion7k with a 7.25\% relative F1-score improvement, and achieve super-human performance on Harm-P with 13.73\% F1-Score improvement. Our approach demonstrates the potential for compact meme classification models, enabling accurate and efficient meme monitoring.
We share our code at \href{jahuerta92/meme-classification-simclip}{https://github.com/jahuerta92/meme-classification-simclip}. \\
\textbf{}
\end{abstract}

\section{Introduction}
For social scientists, memes are ideas spread in a population via imitation~\cite{dawkins2006selfish}. In a similar fashion to a genetic strand, they behave as carriers of concepts and ideas (often extremist) and social values such as equality, racism, and sexism. In social media, memes help spread cultural ideas through images and text, conveying narratives with complex cultural background and context. Social media discourse is guided by memes~\cite{miltner2018internet}, therefore their understanding is crucial. The dissemination of harmful memes can steer entire narratives and harm society at large, as for instance, in the case of coronavirus hoaxes~\cite{de2021internet}. Another popular example is ``Pepe the frog''~\cite{glitsos2019pepe}, which had widespread and neutral connotations in plenty of forums until it was gradually co-opted by alt-right and white supremacist movements~\cite{qu2022evolution} (mainly on the 4chan \texttt{/pol/} community\footnote{\texttt{/pol/} is one of the many forums of the 4chan site, known for their overt racism, antisemitism, among other problematic ideas. We do not include links to the site to avoid encouraging further traffic to it.}), being widely spread during the 2016 US Election. Additionally, it is worth noting that \textit{memes have shifting meaning}.

\begin{figure*}[t]
    \includegraphics[width=\textwidth]{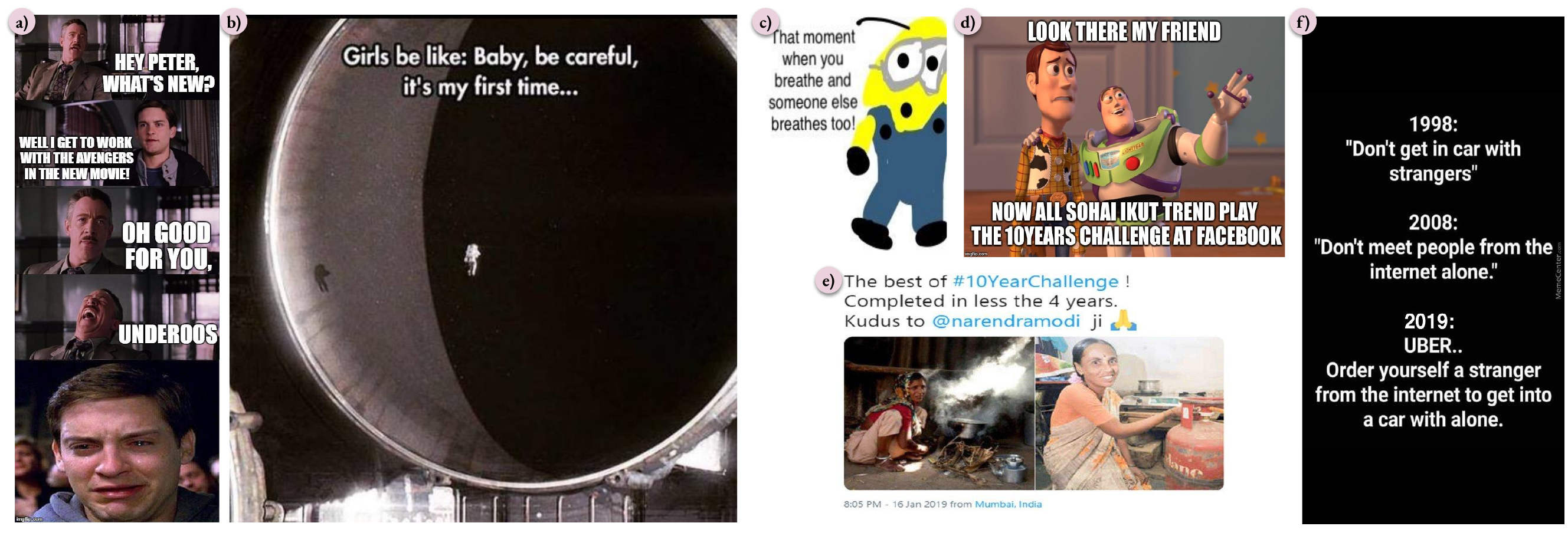}
    \centering \label{fig.memes}
    \caption{Six memes from the Memotion7k~\cite{chhavi2020memotion} training set, illustrating the variety in format. From left to right. \textit{a)} Comic-style template, completely harmless. \textit{b)} Misogynistic image macro, not a common template but still requires context. \textit{c)} Completely ironic, meant to mock similar images. \textit{d)} Image macro, impact text up and down, a common template with contextual meaning. \textit{e)} Just a tweet with an image, still considered a meme.  \textit{f)} Only text available, no image composition or meaning beyond the text. }
\end{figure*}

Another challenge comes with \textit{the inherent multimodality of memes}, as shown in Fig.~\ref{fig.memes}. Take for instance the example presented in Fig.~\ref{fig.memes}(f). The image and the text by themselves are not offensive; however, when put together with added connotations they lead to a highly misogynistic meaning. Last, \textit{memes are inherently subjective}, their interpretation relies on the readers' interpretation of the cross-modal semantics resulting in poor quality annotations in datasets. Quantifying features such as humour, sarcasm, offensiveness, is necessary for meme classification tasks, but at the same time it is an error-prone process, leading to frequent inter-annotator disagreements and label noise.

With these challenges ahead, researchers have proposed several solutions. Leveraging pre-trained models is a popular way to inject knowledge into the model, for example entity, ethnicity and gender detectors. However, the utility of such models is limited by their specific task. For example, it is beneficial to add the ethnicity of the meme subject for hate speech detection, but such a strategy would not be effective for humour classification. Likewise, this approach fails when hate speech detection is performed on images without human subjects. In our study we focus on varied tasks such as humour, sarcasm, offensiveness, hate speech or sentiment analysis, and found out that over-specialization of the model is detrimental to overall meme understanding.

When auxiliary models are excluded, information fusion architectures are popular too, they use pre-trained models to extract textual and visual features and fuse their embeddings to estimate the relationships across modalities. The success of such fusion approaches is usually limited because each model has exclusively unimodal understanding and re-training on low amounts of data is not sufficient to capture the multimodal interactions needed in this particular domain. More recently, some works resort to CLIP embeddings~\cite{radford2021learning}, where the visual and textual modalities are aligned, containing some information about each modality in both branches. The adoption of CLIP has gained momentum in recent literature, showcasing its effectiveness in combining vision and language. 

The interplay between visual and textual cues in memes (which reflects the core of their nature) is well captured by respective representations that lie on a shared vision-language space, and vector operators can then reveal internal associations or contradictions. Using identical projection layers for image and text embeddings preserves this critical condition, which otherwise could be eliminated by the unintended potential creation of semantically different embedding spaces. With this observation at hand, we propose SimCLIP, a Siamese approach, utilizing CLIP features, designed to capture cross-modal interactions effectively for deep meme understanding. The issues of contextual awareness and cross-modal understanding are partially solved at the same time with a large pre-trained Transformer with good off-the-shelf vision-language understanding, thus we consider CLIP for feature extraction. 
The cross-modal feature alignment offered by CLIP, allows to process the embeddings with a Siamese network enabling at the same time (1) cross-modal element-wise operations for interaction-aware representation learning and (2) effective transfer learning to the meme classification domain. 
With simple add-ons and classification heads over CLIP we set the state-of-the-art in some datasets, and achieve on-par performance in others. 
Each dataset poses its own set of challenges and limitations, making multiple benchmarks necessary to demonstrate the effectiveness of our proposal.

\textbf{Contributions}: \begin{itemize}
    \item Treating CLIP as a Siamese network is a novel concept backed by good performance on many datasets.
    \item A novel architecture designed for multimodal meme understanding, exploiting cross-modal interactions with few parameters.
    \item Independence of external knowledge or domain-specific models allows meme analysis by looking exclusively at the image (only OCR is needed).
    \item State-of-the-art performance on meme classification in challenges without resorting to compute-intensive solutions such as ensembles and external knowledge bases. 
    \item Exhaustive validation on seven challenging meme tasks -at the time of writing this article, this is the most extensive evaluation of a meme classification model.
\end{itemize}



\section{Related work} \label{sec:sota} 
Meme classification is a recently introduced task, with challenges and workshops in conferences quickly appearing~\cite{bonheme-grzes-2020-sesam,kiela20hateful,chhavi2020memotion} to stimulate methods that address the detection of hate speech~\cite{sabat2019hate,hermida2023detecting}, misogyny~\cite{fersini2021misogynous}, and even the task of whether an image is a meme or not~\cite{Koutlis2023Jun}. The proposed challenges have contributed to a rich set of benchmarks such as the hateful memes challenge~\cite{kiela20hateful}, which consists of hateful and non-hateful image macros (the most common format for an image meme). Variety is a focus on other works, for example diversity of topic~\cite{Pramanick2021Sep}, where harmful memes are available for two largely different topics: politics and COVID-19. Some tasks rely on small datasets~\cite{suryawanshi-etal-2020-multimodal} for the detection of meme offensiveness. Other works propose composite tasks~\cite{chhavi2020memotion} where many different tasks can be associated with a single meme, e.g., humour, offensiveness, irony, motivation and sentiment. Finally, the detection of disinformation and manipulation through memes, by means of propaganda, has also attracted research interest \cite{dimitrov-etal-2021-detecting}. The meaning of memes arises from images, text and their implicit interactions, finding and detecting these interactions is key to meme analysis and mining tasks.

Incorporating external knowledge in addition to information extracted from the meme's image and text has been proposed by several studies, as a solution to include context to the analysis. For example~\cite{lee2021disentangling} concatenates the text with an entity detector and a demographic detector, enriching the text with context about the target of possible harassment. Also, better predictions can be achieved by using demographic information extracted from the meme~\cite{koutlis2023memefier}. Visual information can also be extracted from regions of interest utilizing pre-trained object detectors~\cite{velioglu2020detecting}. Both approaches can be fused at once, generating captions and extracting region-based visual features~\cite{zhou2021multimodal}. Newer approaches incorporate contextual information by means of knowledge graphs~\cite{kougia2023memegraphs}. 
Also, generating target-aware (e.g. religion) meme representations through GAN techniques has been proposed for zero-shot (in terms of target) hatefulness detection~\cite{zhu2022multimodal}. We highlight MOMENTA~\cite{Pramanick2021Sep} where authors include information about named entities in the text content  
and other textual information such as captions, achieving outstanding performance. Relying on multiple models per modality is the most popular approach to meme understanding as it adds necessary context to the detection task, but carries a steep cost in training and inference. Also we question the ability of these methods to be trained outside of their targeted tasks and datasets, as the added information may be dataset-specific, by relying on dataset-specific domain-exclusive features. This is accented by validation on few datasets despite the current availability of numerous meme-related datasets.  


Instead, a promising direction to explore the interactions of memes in a shared vision-language embedding space, which can be achieved without reliance on ensembles of over-specialized models. For example, token-level cross-modal interactions have been previously studied in MemeFier~\cite{koutlis2023memefier}
and HateCLIPper~\cite{kumar2022hate}. The latter work set a new state-of-the-art performance in a popular benchmark (FBHM) without additional context, by simply merging CLIP representations through a Hadamard product. This powerful approach is reminiscent of how information is merged for Siamese networks, opening up some new paths for research. Seminal works in Siamese networks concatenate all embeddings and their product~\cite{reimers2019sentence} or absolute difference~\cite{chen2017enhanced}. For images, Siamese networks may simply use the product and bias layer~\cite{
zhang2018structured} to merge branches, or auxiliary losses~\cite{liang2019local} on top of the product. Textual Siamese networks apply a similarity function to merge branches~\cite{
reimers2019sentence}. CLIP embeds visual and text data onto the same embedding space, and typical Siamese operators can also work on it by highlighting contradiction and agreement relations, among others. In fact, we find that operators of Siamese networks over CLIP embeddings are under-explored for cross-modal understanding. 


\section{SimCLIP} \label{sec:sclip}
Our method is based on CLIP~\cite{radford2021learning}, which relies on a visual and a textual model, processing each modality to generate an embedding. They are trained with a contrastive objective to maximize the similarity between the associated text-image pairs and minimize similarity between unrelated pairs. At inference time, the CLIP visual encoder (usually a ViT model) produces an embedding of the image in its last projection dense layer; the textual encoder, a minimal Transformer model with small receptive field, performs the same transformation into another embedding. The training method has taught the model to produce aligned embeddings; for example, the embeddings of a duck image and the phrase ``This is a duck'' will have almost maximum similarity. 

Our architecture relies on two fundamental points. First, the creators of CLIP already report results on hate speech detection with reasonable zero-shot accuracy; therefore using the frozen CLIP encoders as feature extractors is a very reasonable choice. In short, thanks to its extensive pre-training, CLIP can already encode information and semantics that are essential to meme understanding~\cite{radford2021learning}. This is likely due to the fact that CLIP has already ``seen'' a fair amount of meme images in the training dataset\footnote{It is impossible to know for certain as CLIP data have not been made public as of yet, but zero-shot performance in the Hateful Memes challenge~\cite{kiela20hateful} are really high at zero-shot without domain adaption, meaning there is a high chance that it has been trained on meme visual and textual language.}. Second, CLIP, despite having two separate independent networks, offers a common embedding space. This means that the features of the image embedding and the text embedding can be jointly interpreted with operators besides concatenation. 
This behaviour, where two embedding types share the same  feature space, resembles the one of a Siamese network. Understanding this similarity is what allows us to leverage CLIP in a Siamese network, as they operate by projecting two pieces of content into the same feature space. This structural compliance allows applying a set of typical Siamese operators onto CLIP projections, in order to preserve feature semantics alignment (consequently allowing for cross-modal operations), while learning the new classification task.

With this approach we can, for example, use common techniques like concatenation of the embeddings and their absolute difference \cite{reimers2019sentence}, or concatenation based on the Hadamard product \cite{chen2017enhanced}. Either technique has shown to be useful in other problems and, 
it should lead to improved performance compared with  simple concatenation. The product and difference of embeddings allows the model to quickly compute interactions between modalities without additional weights and parameters. This efficient comparison of embeddings reduces the need for the network to learn them, cutting down on the trainable parameters. Meme classification datasets are typically small (given the requirements of deep learning), often amounting to fewer than 10k training samples. This limitation requires simplification of the network and the patterns the model needs to learn. Our proposed approach of pre-computing cross-modal interactions is expected to enhance the model's generalization capabilities. Also, in these cases using a large amount of parameters for fine-tuning can be detrimental, thus freezing large parts of the network is also necessary. 

\begin{figure*}[t]
    \includegraphics[width=\textwidth]{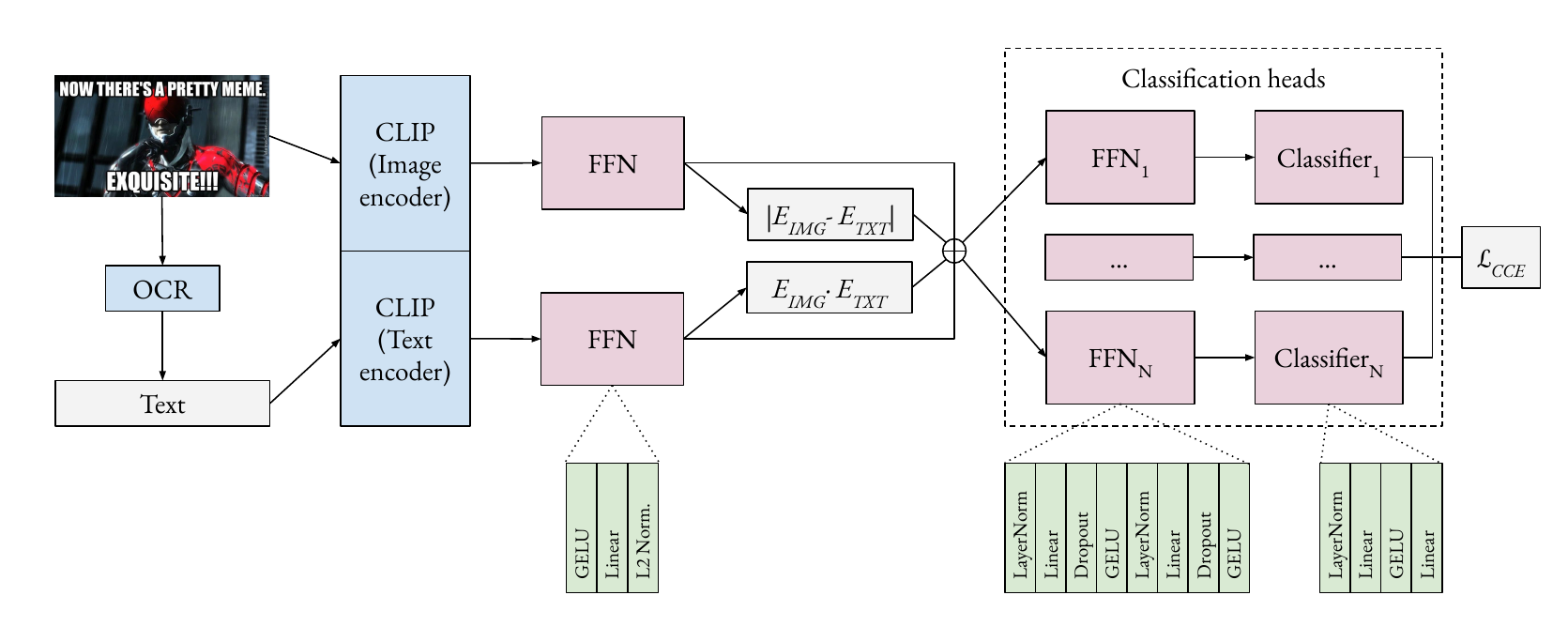}
    \centering 
    \caption{Visualization of SimCLIP architecture. The input image is first processed by an OCR algorithm to extract the text. Image and text are CLIP-encoded and projected using a simple Siamese feed-forward network. Projections are concatenated ($\oplus$) with the absolute difference and Hadamard product to finally be processed by a feed-forward classification head. In multitask settings, losses are summed across heads. In multilabel settings the binary cross-entropy loss is considered, while in multiclass settings we use the categorical cross-entropy loss.}\label{fig:sclip}
\end{figure*}

Our proposed SimCLIP architecture is illustrated in Figure~\ref{fig:sclip}. Each image goes through the model after being preprocessed with an OCR algorithm. Similar works report no improvement or detriment deleting the text from the image~\cite{kumar2022hate}, thus we do not include it in the preprocessing. For preprocessing we simply rescale the image to fit the CLIP input size. However, we add image augmentation after OCR is extracted, applying RandAugment~\cite{cubuk2020randaugment} to memes. This distorts the textual part of the meme but the language information is safely preserved as text thanks to the OCR preprocessing step. Image and text are pairs processed by frozen or partially-thawed CLIP\footnote{Specific CLIP model instance used: \href{openai/clip-vit-large-patch14-336}{openai/clip-vit-large-patch14-336}} encoders, which, after being processed by a shallow network outputs $E_{IMG}$ and $E_{TXT}$ respectively.
The first trainable block of our architecture is the aforementioned shallow network. $E_{IMG}$ and $E_{TXT}$ go through an activation and a linear projection. The projections are L2-normalized for further processing, being transformed to our own feature space. Projections are concatenated along with their absolute difference and Hadamard product as presented in Eq.~\ref{eq.cat}:

\begin{equation} \label{eq.cat}
    E = [E_{TXT}, E_{IMG}, |E_{TXT} - E_{IMG}|, E_{TXT} \cdot E_{IMG}]
\end{equation}

The absolute difference component highlights where the image and text encoder disagree, the higher the disagreement, the larger the activation of the difference, explicitly highlighting contradictory readings of vision and language to the network. If differences represent disagreement, the product will enhance activations on large positive values. If there is high similarity between the projections, the activations of the product will be high in return. Thus, we let the network learn from both representations, their reported differences and similarities.

Some datasets require multiple labels to be predicted at once, so several classification heads are built for each label with $E$ as the input. The classification network has two blocks consisting of normalization, linear layer, dropout and GELU activation. The classifier takes the outputs of the latter to produce a prediction. Each prediction is evaluated with categorical cross-entropy to produce a loss term, balanced with class-weights to account for class imbalance. If there are multiple loss terms, they are summed to produce the final loss $\mathcal{L}_{CCE}$ in case of binary and multi-class problems. In case of multilabel problems, we apply a binary cross-entropy objective $\mathcal{L}_{BCE}$ to a single head to optimize several labels at once. 

\section{Experimental Setup} \label{sec:exp}

\subsection{Datasets}
We use a variety of datasets representing several meme classification tasks. The criteria for selecting the datasets is to contain memes with text (it could be image macros or tweet-type screenshots). We only use the image and the provided text, and we always select the best-quality text extracted that is available in the dataset. A brief description of tasks and size is given in Table~\ref{tab:datasets}, while the details and nuances of each dataset are described in the following points.

\begin{table}
\caption{Dataset summary with task description, training size and eval metrics.}\label{tab:datasets}
\resizebox{\columnwidth}{!}{%
\begin{tabular}{@{}llp{2in}l@{}}
\toprule
\textbf{Dataset} & \textbf{Train size} & \textbf{Task(s)}                                   & \textbf{Metric(s)} \\ \midrule
Memotion7k & 6992 & Task A: Sentiment Analysis, Task C: Fine-grained Offensiveness, Humour, Sarcasm and motivation & Macro F1 \\
FBHM             & 8500                & Hate speech detection                              & AUROC              \\
Harm-C           & 3013                & Harmful content detection (on COVID-related memes) & Accuracy/Macro F1  \\
Harm-P           & 3020                & Harmful content detection (on US Political memes)  & Accuracy/Macro F1  \\
MultiOFF         & 445                 & Offensiveness detection                            & Accuracy/Macro F1  \\
Propaganda       & 950                 & Fine-grained propaganda detection                  & Micro F1/Macro F1  \\ \bottomrule
\end{tabular}%
}
\end{table}

\textbf{Common challenges in meme datasets}: Annotations on all meme datasets are made by small groups of annotators (around three) and majority voting. Therefore, labels may not be entirely reliable, which is a recurring problem. Reliability of labels is not well studied for any of these datasets, but is an important factor that affects training. This is an issue when considering that each dataset has between 1k to 10k examples for training, and datasets such as Propaganda memes and MultiOFF have even less than 1k memes.

The datasets are rather small given the complexity and challenge of the task. Memes are culturally nuanced, can be subtle, overt or sarcastic, and they are sometimes difficult to interpret even for humans due to missing context or other factors. Some datasets derive from challenges that typically distribute the test data only to the participants. Hence, for datasets where test labels are available, we report our metrics on them; otherwise we report results on the validation set.

\textbf{Memotion7k~\cite{chhavi2020memotion}:} This dataset contains 7k memes and presents three tasks. Task A focuses on three-way sentiment analysis, while tasks B and C aim at detecting humour, offensiveness, sarcasm and motivation together. Task B is a binary classification of these different objectives and task C aims to predict their intensity. State of the art results are marginally better than random due to the challenging nature of the inputs and the subjective nature of humour and other sub-objectives. Here, we tackle Tasks A and C. 

\textbf{Hateful Memes Challenge (FBHM)~\cite{kiela20hateful}:} The HMC or FBHM is a dataset by Facebook that sparked a lot of interest in this topic. Despite the interest, the dataset only includes one task, detection of hate in memes. The dataset is very homogeneous as all memes share exactly the same format with little to no variation. The real challenge comes from detecting and attacking the cross-modal offensiveness that appears when text and image are maliciously paired to produce hate.

\textbf{Harm-P \& Harm-C~\cite{Pramanick2021Sep}:} Both datasets stem from the same work, have similar composition of meme types, tackle the exact same task but differ greatly on the topic. Harm-P memes address US Politics while Harm-C is related to COVID-19. Their class labels allow for binary and three-way classification, and we only evaluate using the latter. 

\textbf{MultiOFF~\cite{suryawanshi-etal-2020-multimodal}:} A small dataset that contains memes with offensive language, with the task associated with detecting said offensiveness. The dataset is similar to FBHM in task scope but very different in composition and size, making it much more challenging.

\textbf{Propaganda memes~\cite{dimitrov-etal-2021-detecting}:} The propaganda dataset proposes the classification of propaganda memes, detecting whether they are political propaganda or not. It is as small as MultiOFF. In particular the challenging part of this dataset is the classification of multi-label propaganda-related classes, increasing the difficulty of the problem greatly. It is the only multilabel dataset of the group and Macro F1-score fails to bring meaningful insights into the performance of this model as it is extremely low (less than 15\%). This is the only multi-label problem in this paper.

\subsection{Evaluation protocol}
We employ standard evaluation metrics, namely Macro F1-Score, Accuracy and Area under ROC (AUROC). These are standardized across all datasets as most of them report at least one of them. 
For comparison, we retrieve the best results available in the literature, the baselines on the article and human performance whenever available.

We perform an ablation study to assess the contribution of the embedding combinations, namely the difference, product and concatenation with three alternatives: $E_{CAT} = [E_{TXT}, E_{IMG}]$, $E_{DIFF} = [E_{TXT}, E_{IMG}, |E_{TXT} - E_{IMG}|]$, $E_{PROD} = [E_{TXT}, E_{IMG}, E_{TXT} \cdot E_{IMG}]$. 

\subsection{Implementation details}
Training has been executed on a single NVIDIA GeForce RTX 3090 ti GPU. Each model requires approximately 2 hours to train for the bigger datasets and 10 minutes for the smaller ones. We use the PyTorch\footnote{PyTorch: \href{https://pytorch.org/}{pytorch.org}} backend with the Lightning API\footnote{PyTorch Lightning: \href{https://lightning.ai/docs/pytorch/stable/}{lightning.ai/docs/pytorch/stable/}} to perform and log the experiments. Code for this project is available in: \href{jahuerta92/meme-classification-simclip}{https://github.com/jahuerta92/meme-classification-simclip}.

Training in meme datasets is sensitive to hyper-parameters due to the small amounts of data compared to the complexity of the task. As such we have performed a search for batch size \{64, 32, 16\} and the frozen layers of the transformer \{unfreezing 50\% and 20\% of layers, unfreezing only the CLIP projection layer and freezing CLIP entirely\}. We scale the learning rate to batch size with a set $1e-4$ at 64, $5e-5$ at 32 and $2.5e-5$ at 16. Other parameters are constant, performing learning over 10 epochs, 20\% dropout, 10\% warm-up and constant schedule and a hidden size of 128. In preliminary experiments we found that unfreezing the whole network 1) leads to no improvement in the final model and 2) goes against the premise of the paper of having two models that rely on representations that are bound to the same output embedding space.

\section{Experimental Results}

\subsection{Ablations}
Table~\ref{tab:fbhm} presents the ablations for the Harm, FBHM, MultiOFF and Propaganda datasets. F1-score results are varied, but the full SimCLIP architecture tops the ablations on 2 out of 5 datasets, achieving near top score (within 1\%) in two others. Accuracy is topped in 3 out of 5 datasets and AUROC on 4 out of 5. Overall the full architecture is the best performing one with some special cases; for example there are trade-offs in Harm-C and Harm-P; Accuracy and F1-Score are traded while AUROC remains high. The contribution of the difference seems minimal but in some datasets, it offers a 2\% increase in AUROC, although the real gains of the difference come when combined with the product. Finally, the Propaganda dataset is small and multi-label, and increasing F1-score decreases accuracy and vice-versa while AUROC is within 50\% which is close to random, meaning our model fails on this particular task. 

\begin{table}[]
\centering
\caption{Ablation results on datasets FBHM, Harm-C, Harm-P, MultiOFF and Propaganda.}
\begin{tabular}{@{}ll|rrr@{}}
\toprule
\textit{Dataset} & \textit{Method}     & \multicolumn{1}{l}{\textit{F1 Score}} & \multicolumn{1}{l}{\textit{Accuracy}} & \multicolumn{1}{l}{\textit{AUROC}} \\ \midrule
FBHM             & Concat              & 68.00\%                               & 68.00\%                               & 71.90\%                            \\
                 & Concat + Difference & 66.20\%                               & 66.20\%                               & 71.07\%                            \\
                 & Concat + Product    & \textbf{73.58\%}                      & \textbf{73.60\%}                      & 79.28\%                            \\
                 & Full architecture   & 72.51\%                               & 72.60\%                               & \textbf{79.31\%}                   \\ \midrule
Harm-C           & Concat              & \textbf{56.09\%}                      & 59.74\%                               & 83.21\%                            \\
                 & Concat + Difference & 52.76\%                               & 57.16\%                               & \textbf{85.08\%}                           \\
                 & Concat + Product    & 54.90\%                               & 57.36\%                               & 83.85\%                            \\
                 & Full architecture   & 55.46\%                               & \textbf{60.20\%}                      & 83.82\%                   \\ \midrule
Harm-P           & Concat              & 63.26\%                               & 66.10\%                               & 74.74\%                            \\
                 & Concat + Difference & 63.41\%                               & 64.52\%                               & 76.84\%                            \\
                 & Concat + Product    & 78.25\%                               & \textbf{78.20\%}                      & 90.20\%                            \\
                 & Full architecture   & \textbf{80.01\%}                      & 76.80\%                               & \textbf{91.31\%}                   \\ \midrule
MultiOFF         & Concat              & 64.52\%                               & 64.79\%                               & 67.78\%                            \\
                 & Concat + Difference & 64.71\%                               & 64.71\%                               & 67.69\%                            \\
                 & Concat + Product    & 65.15\%                               & 66.43\%                               & 70.18\%                            \\
                 & Full architecture   & \textbf{66.33\%}                      & \textbf{67.60\%}                      & \textbf{70.39\%}                   \\ \midrule
Propaganda       & Concat              & \textbf{16.69\%}                      & 74.53\%                               & 51.81\%                            \\
                 & Concat + Difference & 15.79\%                               & 80.95\%                               & \textbf{52.12\%}                   \\
                 & Concat + Product    & 13.23\%                               & 83.26\%                               & 49.70\%                            \\
                 & Full architecture   & 11.67\%                               & \textbf{86.65\%}                      & 50.47\%                            \\ \bottomrule
\end{tabular}
\label{tab:fbhm}
\end{table}

\begin{table}[]
\caption{Ablation results on Memotion 7k results on Task 1 and Task 3.}
\resizebox{\columnwidth}{!}{%
\begin{tabular}{@{}ll|rrrr@{}}
\toprule
 &
   &
  \multicolumn{1}{l}{Concat} &
  \multicolumn{1}{l}{Cat.+Diff.} &
  \multicolumn{1}{l}{Cat.+Prod.} &
  \multicolumn{1}{l}{All} \\ \midrule
\textit{\textbf{Task 1}} & F1 Score & 34.68\%          & 35.71\% & 35.29\%          & \textbf{36.92\%} \\
                         & Accuracy & 37.59\%          & 37.24\% & 37.24\%          & \textbf{38.14\%} \\
                         & AUROC    & \textbf{54.81\%} & 54.42\% & 53.80\%          & 53.86\%          \\ \midrule
\textit{Humour}          & F1 Score & 29.21\%          & 27.62\% & 30.09\%          & \textbf{30.81\%} \\
                         & Accuracy & 31.77\%          & 30.22\% & 31.76\%          & \textbf{31.80\%} \\
                         & AUROC    & 55.78\%          & 54.90\% & \textbf{56.57\%} & 56.12\%          \\ \midrule
\textit{Sarcasm}         & F1 Score & 25.03\%          & 26.08\% & 28.08\%          & \textbf{29.51\%} \\
                         & Accuracy & 28.56\%          & 27.26\% & 28.01\%          & \textbf{29.51\%} \\
                         & AUROC    & \textbf{55.56\%} & 53.01\% & 53.38\%          & 54.01\%          \\ \midrule
\textit{Offense}         & F1 Score & 25.42\%          & 26.08\% & 29.19\%          & \textbf{29.37\%} \\
                         & Accuracy & 28.26\%          & 27.88\% & \textbf{32.74\%} & 31.94\%          \\
                         & AUROC    & 55.27\%          & 56.02\% & \textbf{56.13\%} & 55.84\%          \\ \midrule
\textit{Motivation}      & F1 Score & 54.20\%          & 56.60\% & 56.65\%          & \textbf{57.46\%} \\
                         & Accuracy & 54.43\%          & 58.09\% & 56.55\%          & \textbf{57.36\%} \\
                         & AUROC    & 55.79\%          & 58.00\% & 58.84\%          & \textbf{59.80\%} \\ \midrule
\textit{\textbf{Task 3 Avg.}} &
  F1 Score &
  33.46\% &
  34.10\% &
  36.00\% &
  \textbf{36.79\%} \\
                         & Accuracy & 35.76\%          & 35.86\% & 37.27\%          & \textbf{37.65\%} \\
                         & AUROC    & 55.60\%          & 55.48\% & 56.23\%          & \textbf{56.44\%} \\ \bottomrule
\end{tabular}%
}
\label{tab:memotion}
\end{table}

Table~\ref{tab:memotion} includes the ablation results on Memotion7k, in which the proposed model performs best. The difference only improves some scenarios, while the product improves most of them, the full extent of the difference contribution appears when all features are considered. On average we find the full architecture is best for both tasks.


\subsection{Comparison to state of the art}
Table~\ref{tab:comparison} presents the results of the comparative analysis. In FBHM, SimCLIP  falls behind by 3\% compared to the top performing method noting, however, that top models in this dataset are very hard to beat due to extensive use of ensembling beyond CLIP and other pre-processing techniques, including heavy optimization for the challenge. Some novel techniques like MOMENTA, used in other datasets, are outperformed by our model. CLIP is also included in this comparison following the reported results in the original paper; SimCLIP outperforms  CLIP by 2\% in terms of AUROC.

\begin{table}[]
\caption{Comparative results against State-of-the-art models, baselines and human performance (if available)}
\resizebox{\columnwidth}{!}{%
\centering
\begin{tabular}{@{}ll|rll@{}}
\toprule
                    &                                                                            & \multicolumn{1}{l}{F1 Score} & Accuracy                             & AUROC                                \\ \midrule
FBHM                & MemeFier \cite{koutlis2023memefier}                       & \multicolumn{1}{l}{-}        & \multicolumn{1}{r}{73.60\%}          & \multicolumn{1}{r}{80.10\%}          \\
                    & DisMultiHate \cite{lee2021disentangling}                  & \multicolumn{1}{l}{-}        & \multicolumn{1}{r}{75.80\%}          & \multicolumn{1}{r}{82.80\%}          \\
                    & Vilio \cite{muennighoff2020vilio}                         & \multicolumn{1}{l}{-}        & -                                    & \multicolumn{1}{r}{81.56\%}          \\
                    & HateCLIPper \cite{kumar2022hate}                          & \multicolumn{1}{l}{-}        & -                                    & \multicolumn{1}{r}{82.62\%}          \\
                    & MOMENTA \cite{Pramanick2021Sep}                           & \multicolumn{1}{l}{-}        & -                                    & \multicolumn{1}{r}{78.88\%}          \\
                    & Zhu \cite{Zhu2020Dec}                                     & \multicolumn{1}{l}{-}        & \multicolumn{1}{r}{73.40\%}          & \multicolumn{1}{r}{84.60\%}          \\
                    & CLIP \cite{radford2021learning}                           & \multicolumn{1}{l}{-}        & -                                    & \multicolumn{1}{r}{77.30\%}          \\
                    & \textit{SimCLIP}                                                            & \textit{72.51\%}             & \multicolumn{1}{r}{\textit{72.60\%}} & \multicolumn{1}{r}{\textit{79.31\%}} \\ \midrule
Harm-C              & Human \cite{Pramanick2021Sep}                             & 65.10\%                      & \multicolumn{1}{r}{86.10\%}          & -                                    \\
                    & MOMENTA \cite{Pramanick2021Sep}                           & 54.74\%                      & \multicolumn{1}{r}{77.10\%}          & -                                    \\
                    & CLIP \cite{Pramanick2021Sep}                              & 45.55\%                      & \multicolumn{1}{r}{71.05\%}          & -                                    \\
                    & \textit{SimCLIP}                                                            & \textit{55.46\%}             & \multicolumn{1}{r}{\textit{60.20\%}} & \multicolumn{1}{r}{\textit{83.82\%}} \\ \midrule
Harm-P              & Human \cite{Pramanick2021Sep}                             & 70.35\%                      & \multicolumn{1}{r}{92.12\%}          & -                                    \\
                    & MOMENTA \cite{Pramanick2021Sep}                           & 66.66\%                      & \multicolumn{1}{r}{87.14\%}          & -                                    \\
                    & CLIP \cite{Pramanick2021Sep}                              & 60.23\%                      & \multicolumn{1}{r}{80.75\%}          & -                                    \\
                    & \textit{SimCLIP}                                                            & \textit{80.01\%}             & \multicolumn{1}{r}{\textit{76.80\%}} & \multicolumn{1}{r}{\textit{91.31\%}} \\ \midrule
MultiOFF            & MemeFier \cite{koutlis2023memefier}                       & 62.50\%                      & \multicolumn{1}{r}{68.50\%}          & -                                    \\
                    & DisMultiHate \cite{lee2021disentangling}                  & 64.60\%                      & -                                    & -                                    \\
                    & MeBERT \cite{Zhong2022Mar}                                & 67.10\%                      & -                                    & -                                    \\
                    & Best in baselines \cite{suryawanshi-etal-2020-multimodal} & 54.00\%                      & -                                    & -                                    \\
                    & \textit{SimCLIP}                                                            & \textit{66.33\%}             & \multicolumn{1}{r}{\textit{67.60\%}} & \multicolumn{1}{r}{\textit{70.39\%}} \\ \midrule
Propaganda          & VisualBERT COCO \cite{dimitrov-etal-2021-detecting}       & 11.87\%                      & -                                    & -                                    \\
                    & BERT \cite{dimitrov-etal-2021-detecting}                  & 15.68\%                      & -                                    & -                                    \\
                    & \textit{SimCLIP}                                                            & \textit{11.67\%}             & \multicolumn{1}{r}{\textit{86.65\%}} & \multicolumn{1}{r}{\textit{50.47\%}} \\ \midrule
Memotion7K - Task 1 & MemeFier \cite{koutlis2023memefier}                       & 39.60\%                      & -                                    & -                                    \\
                    & George Alexandru et al \cite{vlad-etal-2020-upb}          & 34.50\%                      & -                                    & -                                    \\
                    & Guoym et al \cite{Guo2020Dec}                             & 35.20\%                      & -                                    & -                                    \\
                    & MeBERT \cite{Zhong2022Mar}                                & 37.00\%                      & -                                    & -                                    \\
                    & Best in competition \cite{chhavi2020memotion}             & 35.46\%                      & -                                    &                                      \\
                    & \textit{SimCLIP}                                                            & \textit{36.92\%}             & \multicolumn{1}{r}{\textit{38.14\%}} & \multicolumn{1}{r}{\textit{53.86\%}} \\ \midrule
Memotion7K - Task 3 & MemeFier \cite{koutlis2023memefier}                       & 34.30\%                      &                     -                 &          -                            \\
                    & George Alexandru et al \cite{vlad-etal-2020-upb}          & 31.70\%                      &      -                                &               -                       \\
                    & Guoym et al \cite{Guo2020Dec}                             & 32.30\%                      &    -                                  &                 -                     \\
                    & Best in competition (Avg) \cite{chhavi2020memotion}       & 33.22\%                      &    -                                  &             -                         \\
                    & \textit{SimCLIP}                                                            & \textit{36.79\%}             & \multicolumn{1}{r}{\textit{37.65\%}} & \multicolumn{1}{r}{\textit{56.44\%}} \\ \bottomrule
\end{tabular}
}
\label{tab:comparison}
\end{table}

The Harm-C and Harm-P models are similar in nature but different in their contents and semantics. Our model shows varied performance: in Harm-C we fall far behind human performance on F1-Score and Accuracy; however, our F1-score is the best of the evaluated models. This is explained by our prioritization with class weighting in the training process. Accuracy is penalized by this decision being many points behind MOMENTA. This divide also happens on Harm-P in a similar way, but this time our model proves to be much more powerful. Both MOMENTA and CLIP achieve higher accuracy but much lower F1-score. As a result, the F1-Score of SimCLIP outperforms human scores by 10\% and 13.4\% and 20\% the MOMENTA and CLIP models respectively.

MultiOFF is a small dataset, where we achieve near top performance on available metrics. F1-Score is 1\% less than MeBERT and accuracy is 1\% less than MemeFier. This places SimCLIP near the state of the art performance on this dataset. The DisMultiHate method, which was also in the FBHM evaluation, underperforms in this particular scenario. The small size of the dataset clearly hurts performance for these methods.

In the ablations, we established that SimCLIP could not achieve high performance in the Propaganda dataset leading to near-random results. Surprisingly the performance by VisualBERT COCO is also very low. Multi-modality simply underperforms in this scenario, with BERT achieving better results.

Finally, we take a look at the Memotion 7k tasks. For task 1 we observe that SimCLIP is the third best model (very similar results to MeBERT) in the ranking, outperformed by 2\% by MemeFier. On the other hand for Task 3 we establish a 36.79\% F1-score setting a new state-of-the-art, overcoming MemeFier and other similar models by 2\%.

To summarize our findings, there are two datasets (FBHM and Propaganda) where SimCLIP has exhibited sub-par performance, of which Propaganda seems to be a challenging dataset for all multimodal methods. On the other hand, we set a new standard for F1-Score on Harm-C and Harm-P; in the case of Harm-P, SimCLIP achieves super-human performance. For MultiOFF we achieve near-top performance without compromising either F1-Score or Accuracy. Finally, on Memotion7k we establish a new state-of-the-art performance on Task 3 and achieve fair results in Task 1.
\section{Discussion}
An important issue that our study highlighted is that most of the considered training datasets do not include in-the-wild memes; instead, the samples are specially curated. This means that the real-world generalization capabilities of our framework might still be limited, requiring a much deeper understanding of the complex landscape of Internet memes. To achieve this, a much higher amount of training data would be required due to the complexity of the problem at hand (variety, label quality and context); such amounts of data may exist on social media platforms but does not exist in any structured, high-quality repository. The presented benchmarks of SimCLIP and competing methods over small datasets call into question the generalization capabilities for out-of-domain memes. 

We are aware that original implementations of CLIP has some domain knowledge of memes, but further domain adaptation could be achieved with a larger dataset, which would involve several hundred thousand image-caption pairs. The design of the pre-training task is non-trivial as repeating CLIP pre-training may lead to unforeseen forgetting of previously useful features. Another similar direction would be to design a novel pre-training task for CLIP that could capture cultural features from memes. Other potential solutions could even require capturing the real-world context based on knowledge networks or information retrieval, but this has downsides, as using external models may influence results and bias the resulting model. 

Mislabelling is another challenge that we tried to mitigate but it may have not been enough to increase the variety of data. In the literature, we found several noise handling techniques, all requiring various amounts of fine-tuning. Finding the adequate technique and finding proper hyper-parameters for it, is a challenge that requires extensive trial and error that cannot be automated while varying from dataset to dataset. Techniques such as noise resistant losses and noise adaption layers, are common noise mitigation techniques that require extensive time-consuming experimentation that is beyond the scope of this work. Label noise itself is a very pressing issue in meme analysis due to the subjective nature of humour, offense and harm; that requires further research.

\section{Conclusions} \label{sec:con}
We have studied the problem of meme classification through the use of a CLIP-based architecture. We leveraged Siamese networks as a means to integrate a set base of operators for combining information, including the Hadamard product to highlight similarities in the activations of neurons, while using the absolute point-wise difference to detect discrepancies between models. As CLIP projects its embeddings into a common feature space, these operators remain valid for the CLIP embeddings despite the CLIP weights not being actually identical across modality branches. When we consider CLIP embeddings as the result of the Siamese network we allow for adapted Siamese operators to detect the cross-modal interactions between vision-language features. With this premise, we have been able to match the state-of-the-art in some datasets and set a new one on others, demonstrating good performance across different domains and training setups. At this time this is one of the most exhaustively tested meme models in the literature, including 6 datasets with varied tasks.

We demonstrate the ability of SimCLIP to fuse multimodal information, exploiting the cross-modal interactions with the product and difference of embeddings. The contribution of the product is large while the difference has lesser impact on the final model. Freezing the CLIP network and adding a projection layer is much more efficient than state-of-the-art models, reducing the model size greatly, especially compared to methods based on ensembles of models. 

SimCLIP is self-contained, requiring no external knowledge outside of the meme, achieving competitive results with a far lighter architecture. But this simplicity comes at a cost in some cases. Overall, however, we achieve similar performance to bigger models, sometimes even achieving super-human performance, which makes SimCLIP a versatile and sensible baseline approach for meme analysis.


%
%

\end{document}